%% file: main.tex
\documentclass[sigconf, 10pt, nonacm]{acmart}
\input{0_packages}

\input{0_configs}

\input{0_macros}

\title[]{Blackbox Performance Analysis of LLM-based Network Control}

\author{Mahdi Alizadeh}
\affiliation{\institution{University of Southern California}\country{}
\city{}}

\author{Pooria Namyar}
\affiliation{\institution{Microsoft Research}
\country{}
\city{}}

\author{Arvin Ghavidel}
\affiliation{\institution{University of Southern California}\country{}
\city{}}

\author{Ramesh Govindan}
\affiliation{\institution{University of Southern California}\country{}
\city{}}

\input{tex/abstract}

\begin{document}
\maketitle

\input{tex/intro}
\input{tex/motivation}

\input{tex/design}
\input{tex/multi-stage}
\input{tex/scaling}
\input{tex/constraints}

\input{tex/evaluation}
\input{tex/future_work}
\input{tex/related_work}
\input{tex/conclusion}

\bibliographystyle{ACM-Reference-Format}
\bibliography{ref}

\end{document}

%% file: 0_packages.tex
\usepackage[T1]{fontenc}
\usepackage[utf8]{inputenc}

\usepackage{upquote}

\usepackage{microtype}
\UseMicrotypeSet[protrusion]{basicmath} 

\usepackage{environ}
\usepackage{hyperref}
\usepackage{graphicx,grffile}
\usepackage{multirow}
\usepackage{xspace}
\usepackage{tabularx}
\usepackage{ragged2e}
\usepackage{booktabs}
\usepackage{paralist}
\usepackage[american]{babel}
\usepackage[shortlabels]{enumitem}
\setlist{nosep,leftmargin=*}
\usepackage{courier,color,wrapfig}
\usepackage{balance}
\usepackage{epstopdf}
\usepackage{url}
\usepackage{pifont}
\usepackage{subfigure}
\usepackage[table]{xcolor}

\usepackage{mathtools}
\usepackage{thmtools}
\usepackage[normalem]{ulem}

\usepackage[linesnumbered]{algorithm2e}
\usepackage{etoolbox}
\usepackage[compact,small]{titlesec}
\usepackage{caption}
\usepackage[export]{adjustbox}

\usepackage[capitalize]{cleveref}
\crefformat{section}{§#2#1#3}

\usepackage{tikz}
\usetikzlibrary[shapes.misc, shapes.geometric, shapes.arrows, positioning, decorations.pathmorphing, matrix, graphs, fit, calc, arrows, arrows.meta, backgrounds]

\usepackage{tcolorbox}
\usepackage{csvsimple}

\usepackage[export]{adjustbox}

\usepackage[capitalize]{cleveref}
\crefformat{section}{§#2#1#3}

\usetikzlibrary[shapes.misc, shapes.geometric, shapes.arrows, positioning, decorations.pathmorphing, matrix, graphs, fit, calc, arrows, arrows.meta, backgrounds]

\usepackage{placeins}

\usepackage[colorinlistoftodos,prependcaption]{todonotes}
\usepackage{pifont}

%% file: 0_configs.tex
\renewcommand\footnotetextcopyrightpermission[1]{} %

\PassOptionsToPackage{usenames,dvipsnames}{color} 
\hypersetup{unicode=true,
            colorlinks=true,
            linkcolor=blue,
            citecolor=blue,
            anchorcolor=blue,
            urlcolor=blue,
            breaklinks=true}
\makeatletter
\def\maxwidth{\ifdim\Gin@nat@width>\linewidth\linewidth\else\Gin@nat@width\fi}
\def\maxheight{\ifdim\Gin@nat@height>\textheight\textheight\else\Gin@nat@height\fi}
\makeatother
\setkeys{Gin}{width=\maxwidth,height=\maxheight,keepaspectratio}
\makeatletter
\g@addto@macro{\UrlBreaks}{\UrlOrds}
\makeatother

\newcommand\paraspace{\vspace*{0.25ex}}
\providecommand\parab[1]{\paraspace\noindent\textbf{#1}}

\apptocmd\normalsize{%
\abovedisplayskip=2pt
\abovedisplayshortskip=2pt
\belowdisplayskip=2pt
\belowdisplayshortskip=2pt
}{}{}

%% file: 0_macros.tex
\newcommand{\sysname}[1]{LENS\xspace}

\newcommand{\cmark}{\ding{51}}
\newcommand{\xmark}{\ding{55}}

%% file: tex/abstract.tex
\begin{abstract}
LLMs are becoming an indispensable part of network control systems due to their impressive reasoning. 
  Before such LLM-enabled systems can be deployed, it is imperative to develop tools that can audit their performance on previously unseen inputs.
  However, auditing these systems is challenging due to the inherent opacity of LLMs.
  \sysname{} is the first step towards \textit{black-box} performance analysis of LLM-enabled network control systems.
  It uses novel methods to extend deep reinforcement learning to guide the search for performance gaps and inputs that make LLMs struggle.
  \sysname{} finds large gaps for LLM-based traffic engineering and GPU allocation systems, and even matches existing white-box analyzers designed for hand-crafted TE heuristics.
\end{abstract}

%% file: tex/intro.tex
\section{Introduction}
\label{sec:introduction}

Large language models (LLMs) are moving beyond conversation and code generation to decision-making~\cite{lmte,LLMCoSolver,opro,mllm,selfguidinglm,akts}. 
Motivated by their reasoning and generalization capabilities, research is starting to use them for \textit{network control}: allocating paths, compute, and other resources (\cref{sec:motivation}).
For example, LMTE~\cite{lmte} uses LLMs to solve traffic engineering (TE) problems, exceeding 99.5\% accuracy on held-out demands.

We show that despite this, an adversary without access to a single model weight can construct a \textit{valid} traffic matrix that causes LMTE to leave more than half of a network's traffic unrouted relative to the optimal solution (\cref{sec:evaluation}).
WANs today route terabytes of demand over capacity-constrained links~\cite{soroush,swan,onewan}, and must either drop the resulting unrouted demand, degrading user-perceived performance, or overprovision for it, raising costs.
Thus, before LLM-enabled network control is pervasively deployed, we must address a more challenging issue than average-case accuracy: \emph{how poorly can it perform on valid, realistic inputs, and how would we even find out?}

Prior work on performance analyzers answers exactly this question by searching for \textit{adversarial inputs}, but for a different class of target.
White-box tools require a precise mathematical or logical description of the target~\cite{metaopt,metaopt-nsdi,xplain,formal-network-performance}; gray-box methods require source code, gradients, or access to other system internals~\cite{end-to-end,karimi2026metaease}.
Such tools are effective in a world of hand-written heuristics or solvers whose logic is authored, stable, and inspectable.

LLM-based network control breaks that bargain.
Hosted LLMs expose an API, not weights, prompts, decoding logic, or processing steps.
Even when weights are public, faithfully reducing billions of parameters and their processing logic to a formal specification or source-level model is intractable.
On the other hand, black-box methods such as random search, hill climbing, and simulated annealing need no model of the target and so can be plausibly used for LLM-based network control.
However, they are ineffective because they treat candidate inputs independently or exploit only local feedback, and often fail to uncover meaningful adversarial inputs even for simple heuristics~\cite{metaopt} (\cref{sec:design}).

Given the likely future prevalence of LLM-enabled network control, it is imperative to develop performance auditing techniques that scale with opacity instead of assuming it away.
As more network management tasks employ LLMs~\cite{confucius}, more of their logic will run behind an opaque API rather than in code anyone can read or a model anyone can query for gradients.
A performance analyzer that assumes inspectability will have less and less left to analyze.

\parab{\sysname{}: RL-based search for adversarial inputs.}
In this paper, we introduce \sysname{}, a black-box performance analyzer for LLM-based network control.
\sysname{}, unlike prior work (\cref{tab:prior-applicability}), requires no access to model weights, source code, or other solver internals.
It requires only a representation of valid inputs, an executable interface to the target LLM solver, and a function for scoring the quality of its decisions (\textit{e.g.}, the gap between the LLM's decision and the optimum).

\sysname{} casts adversarial input generation as a reinforcement learning (RL) problem (\cref{sec:design}).
Starting from a valid seed input, the RL policy applies a sequence of changes to the input.
\sysname{} evaluates the modified input, and uses the resulting target-to-optimal performance gap as the reward.
Unlike local black-box search, its search policy reuses experience across evaluations to discover adversarial patterns, such as TE demand configurations that expose bottlenecks consistently mishandled by the target LLM\@.

\parab{Challenges and Contributions.}
Unfortunately, vanilla RL is ineffective for this task for three reasons.

First, the input space is high-dimensional.
Vanilla RL often discovers the coarse shape of an adversarial input but fails to refine its exact values and can get trapped in a local optimum.
\sysname{} addresses this problem with \textit{multi-stage refinement}.
After an initial broad search, it estimates the importance of each input field, freezes the fields responsible for the current gap, and resumes RL over the remaining fields.
Thus, each stage reduces the effective search space and ensures the RL policy focuses on specific parts of the input while preserving earlier progress.

Second, RL requires many reward observations, but computing each reward is expensive: it requires a call to a remote LLM service and an invocation of an optimization solver.
\sysname{} reduces this cost through sparse reward evaluation and result caching.
In addition, it increases throughput with parallel actors and improves efficiency by warm-starting optimization solvers from solutions to nearby inputs.

Third, unconstrained RL exploration can produce unrealistic inputs.
\sysname{} lets operators describe realistic workloads through constraints.
It then restricts the search accordingly, projecting each edit into the feasible space.
The resulting adversarial inputs remain valid and relevant to the target deployment.


\begin{table}[t]
    \centering
    \scriptsize
    \setlength{\tabcolsep}{2pt}
    \begin{tabular}{l c c c}
    \toprule
    \textbf{Tool} &
    \shortstack{\textbf{No Additional}\\\textbf{Formal Modeling}} &
    \shortstack{\textbf{No Source}\\\textbf{Code Access}} &
    \shortstack{\textbf{Structure}\\\textbf{Aware Search}} \\
    \midrule
    \sysname{} (ours) & \cmark & \cmark & \cmark \\
    White-box analysis~\cite{metaopt-nsdi,Virelay,formal-network-performance} & \xmark & \cmark & \cmark \\
    Source code analysis~\cite{karimi2026metaease} & \cmark & \xmark & \cmark \\
    Gray-box analysis~\cite{end-to-end} & \xmark & \cmark & \cmark \\
    Traditional black-box~\cite{davis1991bitclimbing,kirkpatrick1983optimization} & \cmark & \cmark & \xmark \\
    \bottomrule
    \end{tabular}
    \caption{Related work on performance analyzers.}
    \label{tab:prior-applicability}
\end{table}

\parab{Results.}
We evaluate \sysname{} on two LLM-based solvers (\cref{sec:evaluation}).
LMTE~\cite{lmte} achieves more than 99.5\% accuracy on its test set, yet \sysname{} finds traffic matrices that induce a normalized performance gap of more than 50\% without workload constraints and 15.5\% under constraints derived from the training set.
In the constrained setting, \sysname{}'s gap is about 180\% larger than that found by the strongest black-box baseline.
Similarly, for cluster scheduling using LLMCoSolver~\cite{LLMCoSolver},
\sysname{} finds an adversarial gap about 28.5$\times$ larger than the maximum
gap observed on the test set and improves over the strongest black-box baseline
by about 130\%.

Equally importantly, \sysname{} is more general than, or as effective as, some existing white-box analyzers.
For a TE heuristic~\cite{metaopt}, \sysname{} matches the gap found by MetaOpt~\cite{metaopt-nsdi} \textit{without modeling the heuristic at all.}
It also outperforms MetaEase~\cite{karimi2026metaease} by 3.75$\times$ and the strongest traditional black-box baseline by 7.3$\times$ under the same time budget.
\sysname{} generalizes to Teal~\cite{teal}, an RL-based TE solver for which white-box methods (or even gray-box methods such as~\cite{end-to-end}) cannot be applied, and beats the strongest black-box baseline by about 120\%.

\parab{Looking ahead.}
The same property that makes \sysname{} useful today~---~that it needs only an executable interface, not a model~---~ensures that it can scale and adapt as network control systems evolve to use LLMs.
We sketch four directions in \cref{sec:future-work}: scaling to larger problem instances; moving from single high-gap instances to diverse adversarial corpora that can be used for adversarial retraining; repurposing the offline attack policy as an online signal for routing instances to the solver least likely to mishandle them; and, as agentic AI takes on more of network management, using \sysname{} to systematically probe those systems for reliability gaps without ever needing access to their weights or prompts.

%% file: tex/motivation.tex
\section{LLM-based Network Control}
\label{sec:motivation}

In this paper, we consider two LLM-based solvers for network control problems.
Both produce high-quality solutions that are near-optimal on test sets.

\begin{figure}[t]
    \centering
    \includegraphics[width=\columnwidth]{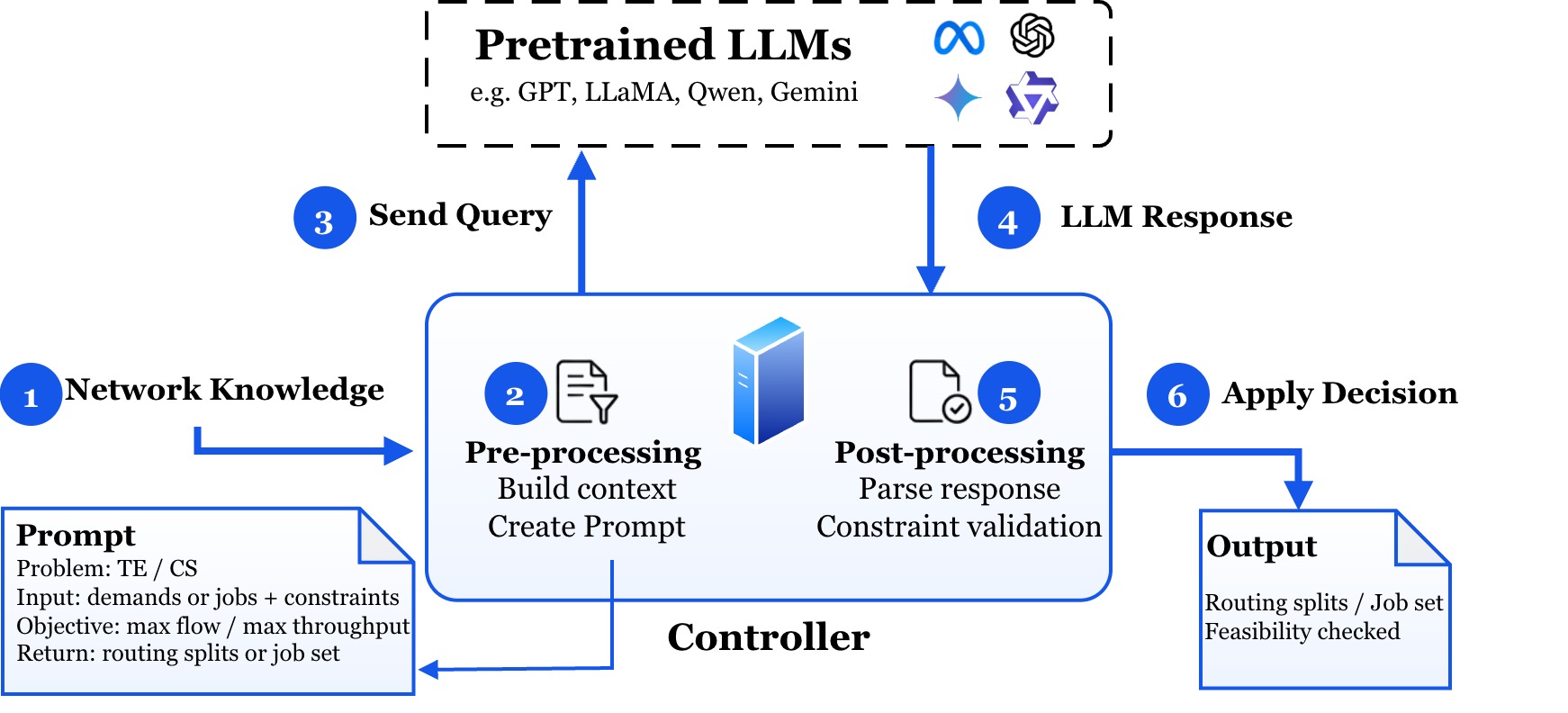}
    \caption{LLM-based traffic engineering workflow.}
    \label{fig:llm-te}
  \end{figure}

\parab{Traffic Engineering.}
TE determines how to split demands between WAN ingress-egress pairs across multiple paths while (a) respecting link capacities, and (b) optimizing a network-wide objective such as maximizing total flow (Max-Flow) or minimizing the maximum link utilization (MLU).
In LMTE~\cite{lmte}, an LLM-based approach to solving TE problems (\cref{fig:llm-te}), a controller collects traffic history and network state, converts them into the model's input format, and sends a query to an LLM\@.
LMTE trains an encoder that outputs an embedding for recent traffic matrices, topology, and link capacities.
It inputs this embedding together with a prompt (\cref{fig:llm-te}) describing the task to a frozen language-model backbone, then applies a lightweight post-processing step to produce split ratios.





\parab{GPU Cluster Scheduling.}
Allocating GPUs to jobs in a cluster while maximizing throughput and respecting capacity constraints can be formulated as a knapsack problem~\cite{copter}.
We have developed an LLM-based solver for this problem, leveraging LLMCoSolver~\cite{LLMCoSolver}, a generic LLM-based optimization solver.
LLMCoSolver must first be fine-tuned on problem instances.
Then, given a cluster configuration and a set of jobs, it returns an allocation of GPUs to jobs.
%




%% file: tex/design.tex
\section{\sysname{}: Black-box RL-based Search}
\label{sec:design}

\parab{Adversarial Input Search.}
Consider the TE problem, and let $Q_{\mathrm{LLM}}(I)$ and $Q_{\mathrm{OPT}}(I)$ denote the quality (\textit{e.g.}, Max-Flow) of the solutions produced by the LLM and the optimal solver on the traffic matrix $I$, respectively.
Given an operator-defined input space $\mathcal{X}$ that captures input constraints, the goal of adversarial input search is to find $I^\star$, the input that maximizes the difference in solution quality: $I^\star \in \arg\max_{I \in \mathcal{X}}
\left[Q_{\mathrm{OPT}}(I)-Q_{\mathrm{LLM}}(I)\right]$.
%
An analogous definition applies to GPU cluster scheduling.

A performance analyzer that searches for adversarial inputs must satisfy two requirements: (a) it must treat the entire LLM-based pipeline as a black box; and (b) it must navigate a high-dimensional, constrained input space effectively enough to uncover meaningful adversarial inputs.

\parab{Exploiting Reinforcement Learning (RL).}
RL learns a policy by interacting with an environment.
At each step, the policy observes the current state, selects an action, transitions to a new state, and receives a scalar reward.
Over repeated interactions, the policy learns which actions are likely to lead to a high-reward state.
\begin{table}[t]
\centering
\small
\begin{tabularx}{\columnwidth}{l X X}
\toprule
 & \textbf{TE} & \textbf{GPU Scheduling} \\
\midrule
\textbf{State} & Demand matrix of src–dest traffic volumes. & Job matrix of throughput and GPU requirements. \\ \hline
\textbf{Action} & Edit multiple demand entries. & Edit throughput/GPU use of multiple jobs. \\ \hline
\textbf{Reward} & Routed flow difference (OPT vs. LLM). & Selected throughput difference (OPT vs. LLM). \\
\bottomrule
\end{tabularx}
\caption{Mapping RL to specific problem domains.}
\label{tab:mdp_mapping}
\end{table}

Adversarial input search can be modeled using RL (\cref{tab:mdp_mapping}).
The current candidate input represents the state, an edit to that input constitutes the action, and the performance gap between the LLM-based system and the optimal solver defines the reward.

This approach satisfies our two requirements.
In RL, the search policy does not need the LLM source code, model weights, or gradients.
It can also search efficiently; rather than treating each candidate input as an isolated trial, the policy learns from prior inputs and rewards to propose promising edits that quickly lead to adversarial inputs.



\parab{Challenges.}
Despite this natural fit, we find that off-the-shelf vanilla RL is sub-optimal for \sysname{}'s adversarial input search.
First, the input space is high-dimensional, which hinders exploration; as a result, vanilla RL often fails to find strong adversarial inputs and can become trapped in local optima (\cref{sec:multi-stage-training}).
Second, RL learns through many multi-step trials, but evaluating the reward at each step is expensive because it requires invoking a remote LLM service and an optimization solver (\cref{sec:runtime-scaling}).
Third, unconstrained exploration can produce invalid or implausible inputs irrelevant to deployment (\cref{sec:constraint-application}).


\subsection{Architecture Overview}
\label{sec:architecture-overview}


\parab{Deep RL learner} observes the current input $I_t$ and proposes an edit $a_t$ intended to increase the performance gap.
During search, it updates its policy from observed transitions $(I_t,a_t,I_{t+1},r_t)$ so that edits that repeatedly produce larger gaps become more likely.
Since the learner uses only this input-output feedback, it does not require access to the internal operation of the LLM-based system.

For training the learner's policy, \sysname{} uses Soft Actor-Critic (SAC)~\cite{sac}.
Compared with an on-policy learner such as PPO~\cite{ppo}, this off-policy design better fits our setting for two reasons.
First, SAC's entropy objective encourages continued exploration, helping the policy avoid local optima.
Second, SAC stores transitions in a replay buffer and reuses them across policy updates, whereas on-policy methods discard them after each update.
This reuse allows the policy to learn complex adversarial patterns from past experience.

\begin{figure}[t]
    \centering
    \includegraphics[width=\columnwidth]{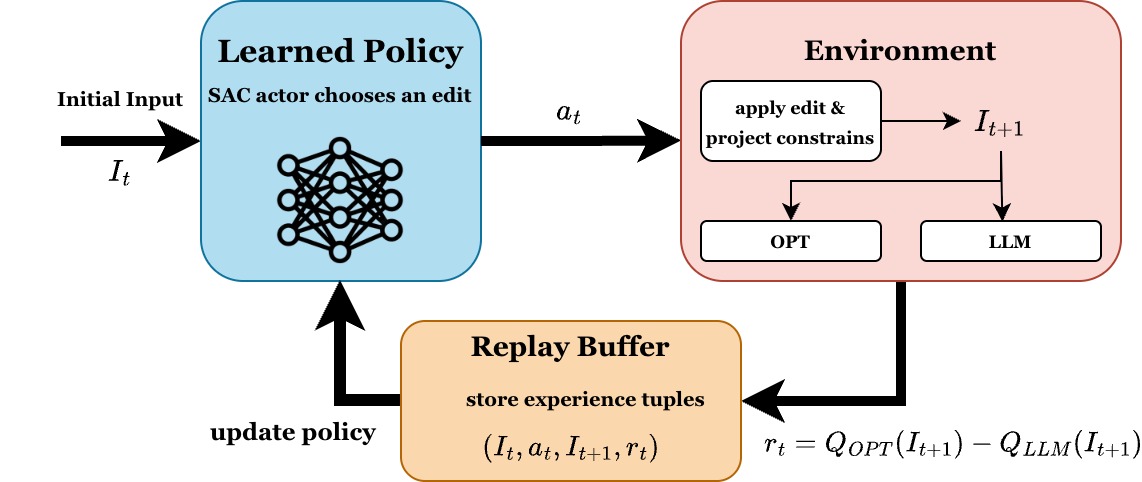}
    \caption{\sysname{} architecture. The policy edits a structured input, the
    environment projects the edit back into the valid input space, and the
    $Q_{\mathrm{OPT}}(I)$ and $Q_{\mathrm{LLM}}(I)$ executors evaluate the
    result. The measured gap is stored as a reward and reused by the learner.}
    \label{fig:metarl-design}
\end{figure}

The policy's feature extractor matches the input structure.
We use dense graph convolutions for TE demand matrices and an MLP for vector-valued cluster scheduling inputs.

\parab{Environment} maintains the current input and applies the edit proposed by the learner.
It then calls the LLM-based solver and the optimization solver to obtain their solutions, computes the performance gap
$Q_{\mathrm{OPT}}(I_{t+1})-Q_{\mathrm{LLM}}(I_{t+1})$, and stores the observed transition and reward in the replay buffer.

In traffic engineering, an edit may increase or decrease the demands between multiple source--destination pairs.
In cluster scheduling, it may change the priority, arrival time, duration, or resource requirements of multiple jobs.

\parab{Replay Buffer} stores transitions and rewards $(I_t,a_t,I_{t+1},r_t)$ generated during the search.
The policy network reuses these transitions to update its parameters frequently, which allows it to learn complex adversarial patterns from past experience and prevents it from getting stuck in local optima~\cite{sac}.

%% file: tex/multi-stage.tex
\subsection{Multi-Stage Refinement}
\label{sec:multi-stage-training}

Vanilla RL often learns the coarse-grained structure of an adversarial input but struggles to refine it into a high-impact adversarial input.
In TE, for example, the policy may learn which demands should increase or decrease to pressure particular bottlenecks, but may fail to identify their precise values.
Two factors make this refinement task challenging for RL\@.
First, the input space is high-dimensional, and the performance gap landscape is non-convex.
Because each action can modify many fields (\textit{e.g.}, traffic demands, or job GPU allocations), exploring a better value for one field can undo useful assignments elsewhere.
Second, the scalar reward conveys only the aggregate gap and does not attribute the gap to specific fields.
Using vanilla RL produces gaps $3$--$5\times$ smaller than those identified by \sysname{} (\cref{sec:eval-ablations}).

\sysname{} overcomes this limitation with multi-stage refinement.
\sysname{} first uses RL to discover a coarse adversarial pattern.
It then identifies the fields most responsible for the current gap, freezes their values, and resumes refining the remaining fields.
Repeating this process preserves progress from earlier stages while reducing the number of fields that remain subject to exploration.
The learner can therefore shift from discovering the overall adversarial pattern to refining the values of the remaining unconstrained fields.

To identify critical fields, \sysname{} assigns an importance score to each field in the current input.
The score measures how strongly the current gap is sensitive to that field's value.
%
%
Formally, let $I=(v_1,\ldots,v_n)$ be the current adversarial input and
$R(I)=Q_{\mathrm{OPT}}(I)-Q_{\mathrm{LLM}}(I)$ be its performance gap.
For each field $i$, \sysname{} samples $M$ valid alternative values, substitutes each one while holding all other fields fixed, and evaluates the resulting gaps $R_{i,1},\ldots,R_{i,M}$.
The importance score for field $i$ is defined as $R(I) - \frac{1}{M}\sum_{j=1}^{M} R_{i,j}$.
%
%
A large positive score means that perturbing the field typically reduces the
performance gap, so \sysname{} freezes its current value in subsequent stages.
Fields with scores near or below zero remain open to optimization by the RL policy.
\sysname{} then repeats the search-and-freeze process until it exhausts its stage budget.

%% file: tex/scaling.tex
\subsection{Scaling \sysname{}}
\label{sec:runtime-scaling}

In theory, after every proposed edit the RL learner must compute a new performance gap as its reward.
Computing this gap can dominate runtime: (1) the target may be a cloud-hosted LLM service, and (2) computing the optimal solution can be extremely slow.
\sysname{} combines three techniques to accelerate RL search.
These optimizations have minimal impact on the quality of the discovered inputs (\cref{sec:eval-ablations}).

\parab{Sparse rewards.}
Because each edit is bounded, the performance gap often changes slowly between consecutive inputs.
\sysname{} therefore evaluates the performance gap every $k$ steps, or sooner if the cumulative distance from the last evaluated input exceeds a configured threshold.
At intermediate steps, the policy receives the most recently measured gap as its reward.
This avoids redundant LLM and solver calls while retaining trajectory-level feedback.

\parab{Reward caching and warm starts.}
\sysname{} caches each evaluated input and its measured gap.
When an identical input reappears under the same solver configuration, \sysname{} reuses the cached gap without re-evaluating the target LLM or the optimization solver.
For nearby inputs, it uses the cached optimal solution to warm-start the optimization solver.
The optimizer still computes the solution for the new input, but typically converges faster.
Together, caching and warm starts minimize redundant computation and reduce latency.

\parab{Parallel actors.}
\sysname{} runs independent search instances in parallel from different seed inputs to increase search coverage.
The actors explore different regions concurrently and periodically share high-gap candidates or a replay buffer.
%

%% file: tex/constraints.tex
\subsection{Enforcing Constraints on Inputs}
\label{sec:constraint-application}



Operators typically seek adversarial inputs consistent with their expected workloads.
\sysname{} allows them to define an admissible search space $\mathcal{X}$ that captures both validity and the operating conditions relevant to their deployment, then searches only for adversarial inputs within $\mathcal{X}$.
For example, since WAN traffic matrices are often sparse and local~\cite{metaopt-nsdi,262026}, an operator may constrain demand density and source--destination locality when evaluating LLM-based TE.

\sysname{} enforces these requirements through a projection operator, loosely inspired by the projection step in projected gradient descent (PGD~\cite{boyd_co}).
After each edit, it projects the candidate to the nearest feasible input in $\mathcal{X}$, evaluates only the projected input, and uses that input as the state for the next RL step.
We have implemented projection operators for constraints that specify bounds on individual input fields or aggregate quantities, as well as sparsity and locality.
For example, for locality, \sysname{} sets demands between source--destination pairs that exceed a certain distance threshold to zero.
For density, it retains the allowed number of positive demands and sets to zero the smallest remaining entries.
Adding a new type of constraint requires only implementing its projection operator; the LLM-based system and optimal solver remain black boxes.

%% file: tex/evaluation.tex
\section{Evaluation}
\label{sec:evaluation}

We evaluate \sysname{} on two LLM-based optimization solvers: LMTE~\cite{lmte} for traffic engineering and LLMCoSolver~\cite{LLMCoSolver} for cluster scheduling (\cref{sec:motivation}).
We show that \sysname{} can expose large performance gaps without access to their internal solver pipelines.
We then use \sysname{} to analyze a heuristic and a non-LLM learning-enabled system to demonstrate its generality and effectiveness.

\parab{Experimental setting.}
We run \sysname{} and each baseline with five different random seeds and report the largest gap found across these runs.
All methods are configured to utilize all available CPU threads.

\begin{figure}[t]
	\centering
	\includegraphics[width=\columnwidth]{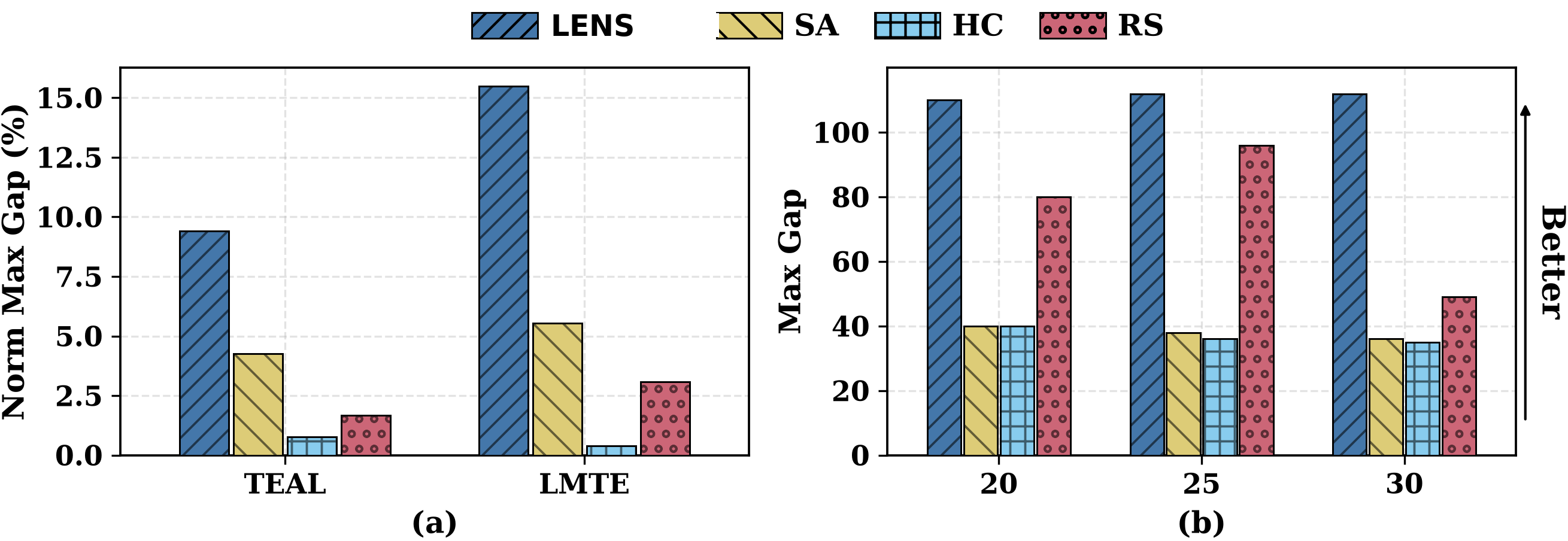}
	\caption{\sysname{} outperforms black-box baselines for (a)
		traffic engineering targets, LMTE and Teal, and (b) cluster scheduling with
		LLMCoSolver. Teal is the learned target discussed in \cref{sec:eval-general-targets}.}
	\label{fig:eval-llm}
\end{figure}

\parab{Baselines.}
We compare \sysname{} with generic black-box search methods: random search, simulated annealing, and hill climbing~\cite{kirkpatrick1983optimization,davis1991bitclimbing}.
We also compare against MetaOpt~\cite{metaopt-nsdi}, a white-box analyzer that computes a provably optimal gap, and MetaEase~\cite{karimi2026metaease}, a source-code-based analyzer.

\subsection{LLM-Based Network Control}
\label{sec:eval-llm}
\label{sec:eval-te}
\label{sec:eval-resource}

\parab{\sysname{} exposes a 57.51\% normalized gap in LMTE.}
We measure the difference between the traffic routed by the optimal solver and LMTE, normalized by the topology's total link capacity.
We evaluate LMTE on B4~\cite{B4} and precompute $K=4$ shortest paths for every source--destination pair.
We use Qwen~2.5 8B as LMTE's backend LM and fine-tune LMTE's embedding layers on 20K gravity-model traffic matrices until its accuracy exceeds 99.5\% on held-out inputs.

Without workload constraints, \sysname{} finds a normalized gap of 57.51\%.
We then restrict the search using bounds derived from the training set: individual demands may not exceed the largest observed demand, and total demand may
not exceed the maximum observed aggregate volume.
Even so, \sysname{} finds a 15.4\% normalized gap.

\parab{\sysname{} exposes a gap over $25\times$ larger than LLMCoSolver's test-set gap.}
Following~\cite{copter}, we formulate cluster scheduling as a 0/1 knapsack problem with a capacity of 200 GPUs; each job requests between 1 and 40 GPUs and has an associated throughput.
We measure the difference between the total throughput achieved by the optimal solver and LLMCoSolver.
We use Qwen~2.5 8B as the backend LM and fine-tune it on 500K randomly generated instances.

The largest gap on the test set is 4 throughput units.
We constrain \sysname{}'s search using the normalized nearest-neighbor distance from the adversarial input to the training set so that generated instances remain close to the training data.
Even under this constraint, \sysname{} finds inputs with a gap exceeding 100 throughput units---more than $25\times$ the largest test-set gap.





\parab{\sysname{} outperforms all black-box baselines (\cref{fig:eval-llm}).}
%
%
For LMTE, \sysname{} finds a gap that is 180\% larger than that found by the strongest black-box baseline.
For LLMCoSolver, \sysname{} finds the largest gap at every evaluated problem size, and the difference grows as the number of jobs increases.

\subsection{Beyond LLM-based Network Control}
\label{sec:eval-general-targets}


\parab{\sysname{} matches or exceeds the white-box analyzers for Demand Pinning.}
Demand pinning (DP)~\cite{278322,metaopt} routes demands that fall below a specified threshold along their shortest paths and solves an optimization problem for the remaining large demands.
We set its threshold to 5\% of the average link capacity and bound each demand by half of the average link capacity.
We use B4 and $K=4$ shortest paths.

\begin{figure}[t]
	\centering
	\includegraphics[width=\columnwidth]{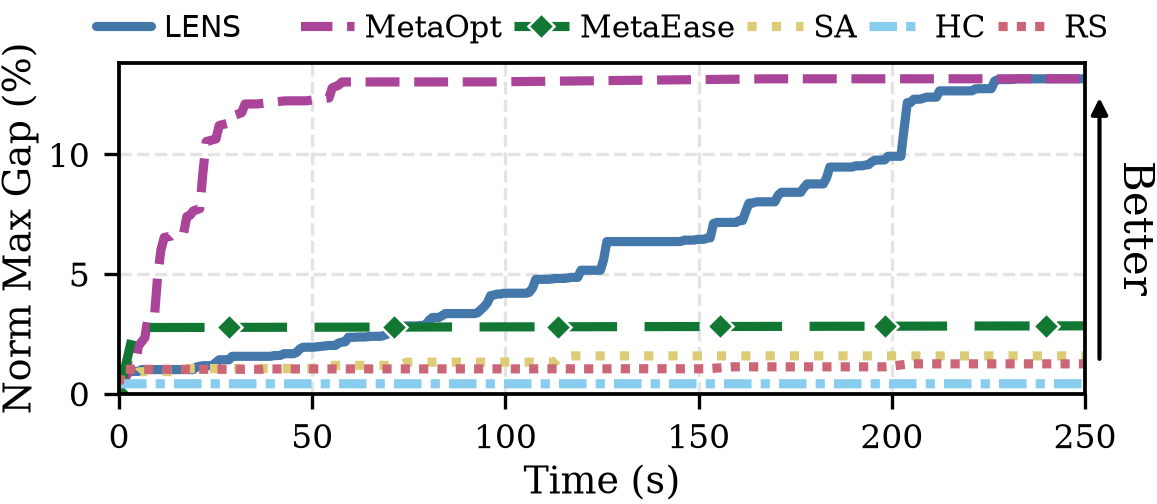}
	\caption{\sysname{} matches MetaOpt, while outperforming MetaEase and black-box
		baselines under the same time budget.}
	\label{fig:eval-te-dp}
\end{figure}

Both MetaOpt and MetaEase can analyze DP.
\cref{fig:eval-te-dp} shows that \sysname{} eventually matches MetaOpt's gap while outperforming MetaEase by $3.75\times$ and the strongest black-box baseline by $7.3\times$.
MetaOpt uses a mathematical model of the heuristic to provably compute the worst-case gap.
In contrast, \sysname{} requires only black-box input-output access to DP and still reaches the same gap, albeit at a slower rate because it lacks access to DP's internals.


\parab{\sysname{} also applies to non-LLM learned systems.}
Teal~\cite{teal} is an RL-based TE system that uses a graph neural network to predict traffic splits and ADMM~\cite{boyd_admm} to repair infeasible outputs.
This combination of a neural policy and iterative repair cannot be formulated in MetaOpt~\cite{metaopt-nsdi} or analyzed by end-to-end gradient analysis~\cite{end-to-end}.

\sysname{} treats Teal as a black box and can analyze its performance.
We train Teal's open-source implementation on the same dataset used for LMTE, and the resulting model achieves more than 99.5\% accuracy on the test set.
Under input constraints, \sysname{} finds a normalized gap of 9.4\%, outperforming all black-box baselines.
Without these constraints, \sysname{} finds a gap of 64.70\%.

\subsection{\sysname{}'s design choices}
\label{sec:eval-ablations}

\begin{figure}[t]
	\centering
	\includegraphics[width=\columnwidth]{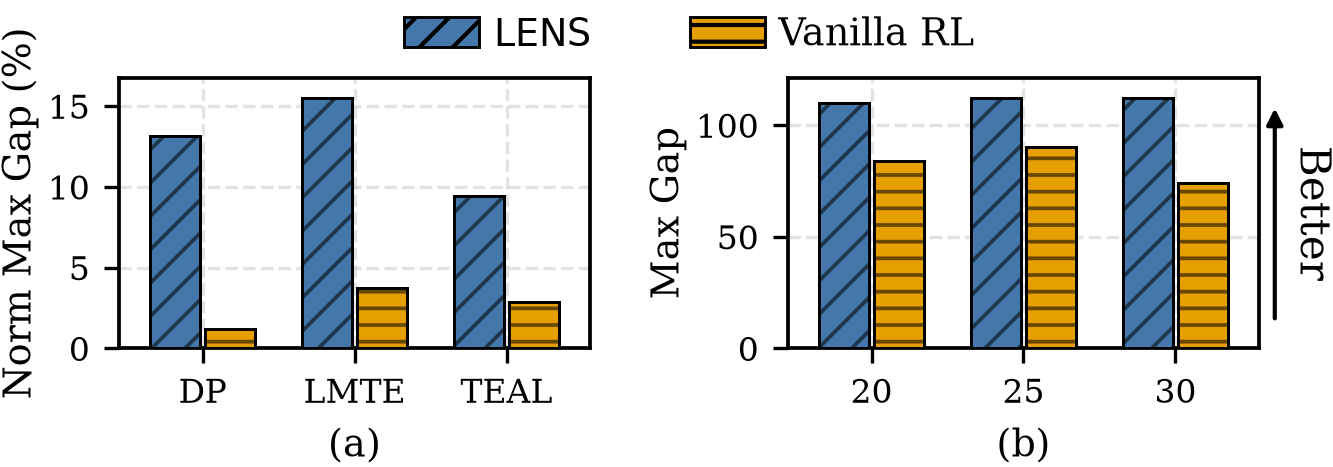}
	\caption{\sysname{}'s multi-stage refinement finds larger gaps than
		vanilla RL under the same search setup. (a) Traffic engineering, (b) Cluster scheduling with different job counts.}
	\label{fig:eval-metarl-vs-vanillarl}
\end{figure}

\parab{Multi-stage refinement.}
Figure~\ref{fig:eval-metarl-vs-vanillarl} compares \sysname{} to vanilla RL under the same policy and search budget.
Multi-stage search finds $50 - 1000 \%$ larger gaps on both TE and cluster scheduling problems.

\begin{figure}[t]
	\centering
	\includegraphics[width=\columnwidth]{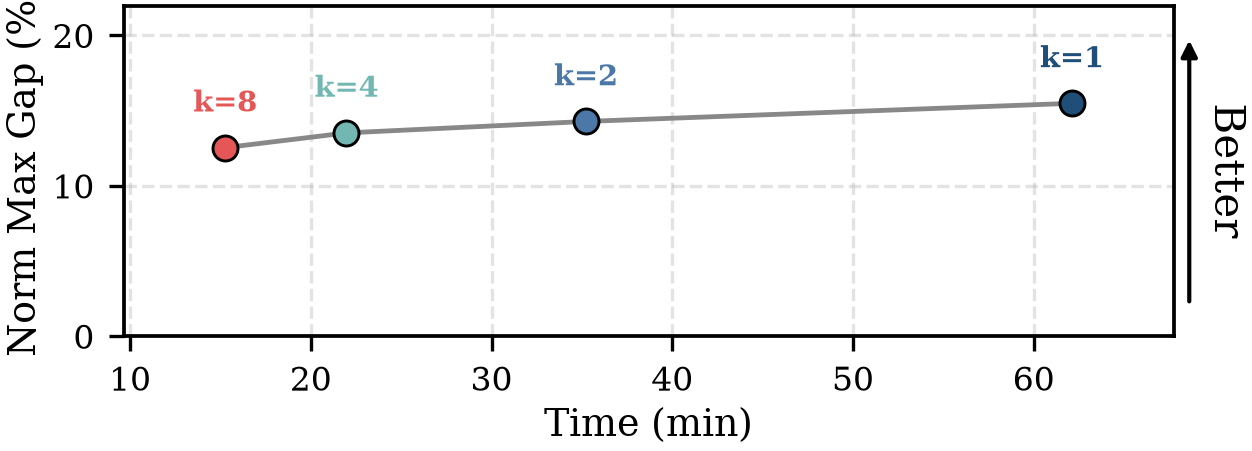}
	\caption{Sparse reward frequency $k$ trades search quality for runtime on
		LMTE.}
	\label{fig:eval-reward-freq}
\end{figure}

\parab{Sparse rewards.}
\sysname{} computes the reward once every $k$ steps (\cref{sec:runtime-scaling}).
Figure~\ref{fig:eval-reward-freq} depicts the discovered gap and the runtime for different values of $k$.
Increasing the step size from $k{=}1$ to $k{=}2$ yields an $8\%$ reduction in the discovered gap but provides a $1.5\times$ speedup.
For $k=4$, these numbers are $13\%$ and $2.8\times$, respectively.

\parab{Warm starts.}
%
We have compared the runtime of \sysname{} with and without warm starting for both LMTE and DP (graphs omitted).
For DP, where Gurobi optimization calls dominate the execution, warm-starting yields a substantial $3.1\times$ speedup at $k{=}1$ (reducing runtime from $23.6$ to $7.5$ minutes).
Conversely, LMTE exhibits a negligible $3\%$ speedup because its runtime is dominated by LLM query latency rather than the optimization solver.

%% file: tex/future_work.tex
\section{Future Work}
\label{sec:future-work}

\parab{Broader and larger-scale domains.}
\sysname{} is a black-box framework, so we can evaluate it on a broader set of problems.
New domains include packet scheduling~\cite{PACKS} and caching~\cite{caching}.
To measure adversarial effectiveness and scalability, we can evaluate over larger WAN topologies for TE or more diverse cluster job sets.
This would likely unearth opportunities for improving \sysname{}'s efficiency and effectiveness.

\parab{Improving LLM-solver robustness.}
\sysname{} currently focuses on finding high-gap instances during search.
A natural extension is to generate a diverse corpus of adversarial inputs, including inputs constrained to a target workload distribution.
These corpora can help increase the robustness of learned solvers through adversarial retraining and stress-test suites, with explicit tracking of the trade-off between worst-case improvement and average-case performance.

\parab{From attack policy to online solver selection.}
The policy learned by \sysname{} maps the given input to an edit that increases the performance gap.
The same representation can be reused in production as a routing signal to select among a portfolio of solvers.
Given an incoming instance, a selector can predict which heuristic or solver is likely to have the smallest expected gap and dispatch accordingly.
This turns \sysname{} from an offline analyzer into an online decision layer for adaptive solver orchestration.

\parab{Generalizing to other agentic AI systems.}
Agentic AI is increasingly used in network management~\cite{confucius}.
A promising direction is to define domain-specific performance metrics for these agents and use \sysname{} to search for adversarial inputs that expose reliability gaps.
Because \sysname{} only needs a black-box execution interface, it can evaluate and stress-test agentic systems across many domains without requiring access to internals such as model weights or prompts.

%% file: tex/related_work.tex
\section{Related Work}
\label{sec:related-work}

\parab{Performance analyzers.}
MetaOpt~\cite{metaopt,metaopt-nsdi}, XPlain~\cite{xplain}, and other domain-customized analyzers~\cite{ccac,formal-network-performance,Virelay} require users to formally specify the target and constraints in a specific language.
MetaEase~\cite{karimi2026metaease} reduces this modeling burden by analyzing
the source code.
These methods remain inapplicable when the complete solver is available only
through an API. 
In contrast, \sysname{} only requires an executable target and a performance scoring function (\cref{tab:prior-applicability}).

\parab{Black-box search.}
Random search, hill climbing~\cite{davis1991bitclimbing}, and simulated
annealing~\cite{kirkpatrick1983optimization} are general.
However, they rely on local feedback and can be trapped in local optima. They often fail to find useful adversarial inputs even for simple heuristics (\cref{fig:eval-te-dp}).

\parab{Robustness of learned systems.}
Adversarial ML methods attack neural policies, image classifiers, and graph
neural networks~\cite{huang2017adversarial,kurakin2018adversarial,ma2020practical}.
These methods target classification or policy outputs and often assume model or gradient access, rather than gaps relative to an optimum.
\cite{gilad19robustifying} uses RL to generate adversarial inputs for RL-based network
controllers, but does not address high-dimensional, constrained optimization instances.
\sysname{} adds multi-stage refinement to make RL effective in this setting.

\parab{Robustness of LLMs.}
LLMs are increasingly used for optimization and network management~\cite{lmte,LLMCoSolver,opro,dragon_llm,architects,confucius,llm_network_config,llm_evolutionary}.
Existing evaluations report average accuracy, feasibility, or task success, while LLM red-teaming benchmarks~\cite{liang2023holistic,harmbench,jailbreakbench,llm_worstcase,netarena}
focus on unsafe generations or agent behavior.
\sysname{} instead searches for valid inputs where the LLM's decision is substantially worse than the optimal.

%% file: tex/conclusion.tex
\section{Conclusion}
\label{sec:conclusion}

\sysname{} is a black-box analyzer for finding adversarial inputs that widen the gap between a target LLM-based solver and an optimal baseline.
It applies to heuristics, learned pipelines, LLM solvers, and agentic AI systems without requiring models, source code, or gradients.
For traffic engineering and cluster scheduling, \sysname{} matches or exceeds prior analyzers on classical targets, outperforms black-box local search, and exposes large gaps in high-accuracy learned and LLM solvers that average-case benchmarks miss.
As such, \sysname{} catalyzes the exploration of performance analyzers that improve the auditability of opaque agentic systems.